\documentclass[runningheads]{llncs}

\pdfoutput=1
\usepackage{hyperref}
\hypersetup{
    pdftitle={SciRecSys: A Recommendation System for Scientific Publication by Discovering Keyword Relationships}, 
    pdfauthor={Vu Le Anh, Hai Vo Hoang, Hung Nghiep Tran, Jason J. Jung}, 
    pdfsubject={Computer Science}, 
    pdfkeywords={Artificial Intelligence, Machine Learning, Data Mining, Markov Chain Monte Carlo, Random Walks, Bibliographic Databases, Recommendation Systems, Keyword Ranking}, 
    pdffitwindow=false,     
    pdfstartview={FitH},    
    pdfcreator={Hung Nghiep Tran}, 
    pdfproducer={Hung Nghiep Tran} 
}

\usepackage{tabularx}
\usepackage{multirow}
\usepackage{hhline}
\usepackage{verbatim} 
\usepackage{subfig}
\usepackage{times}
\usepackage{graphicx}
\usepackage{makeidx,latexsym,amssymb}  
\newcolumntype{R}[1]{>{\raggedright\arraybackslash}p{#1}} 
\newcolumntype{C}[1]{>{\centering\arraybackslash}p{#1}} 
\newcolumntype{L}[1]{>{\raggedleft\arraybackslash}p{#1}} 
\begin{document}
\mainmatter              
\title{SciRecSys: a Recommendation System for Scientific Publication by Discovering Keyword Relationships}
\titlerunning{SciRecSys: a Recommendation System for Scientific Publication}  
\author{Vu Le Anh\inst{1,2}\and Hai Vo Hoang\inst{3}\thanks{Corresponding Author: Hai Vo Hoang, Information Technology College, Ho Chi Minh, Vietnam. Phone/Fax: +84 938 642 717. Email: vohoanghai2@gmail.com.}\and Hung Nghiep Tran\inst{4}\and
Jason J. Jung\inst{5}}
\authorrunning{Vu et al.}   
%
%
\institute{Nguyen Tat Thanh University, Ho Chi Minh city,  Vietnam\\
\and Big IoT BK Project Team, Yeungnam University, Gyeongsan, Korea\\
\and Information Technology College, Ho Chi Minh city, Vietnam\\
\and University of Information Technology, Ho Chi Minh city, Vietnam\\
\and Chung-Ang University, Seoul, Korea\\
\email{lavu@ntt.edu.vn, vohoanghai2@gmail.com, nghiepth@uit.edu.vn, j2jung@gmail.com}
}

\maketitle

\begin{abstract}
In this work, we propose a new approach for  
discovering various relationships among keywords over the scientific publications based on a Markov Chain model. It is an important problem since keywords are the basic elements for representing abstract objects such as documents, user profiles, topics and many things else. Our model is very effective since it combines four important factors in scientific publications: content, publicity, impact and randomness. Particularly, a recommendation system (called SciRecSys) has been presented to support users to efficiently find out relevant articles.

\end{abstract}
\section{Introduction}
Keyword-based search engines (Google, Bing Search, and Yahoo) have emerged and dominated the Internet. The success of these search engines are based on the study of keyword relationships and keyword indexes. The task of measuring keywords' relationships is the basic operation for building the related network of abstract objects which are applied in many problems and applications, such as document clustering, synonym extraction, plagiarism detection problem, taxonomy, search engine optimization, recommendation system, etc. 
In this work, we focus on two problems. First, we study the rank, inference and similarity of keywords over scientific publications on assuming that the keywords belong to a \textit{virtual ontology of keywords}. The inference relationship will help us find \textit{parents, children}, the similarity relationship will help us find \textit{siblings} of a given keyword and the rank of keywords will determine how important they are. Second, we apply these relationship in our scientific recommendation system, SciRecSys.
The first problem is not easy since we have to consider four main factors:  
\begin{itemize}
\item[i-]~~\textit{Content factor}. The meaning of the keyword should be considered in the context of its paper. The paper itself is determined by its keywords; 
\item[ii-]~~\textit{Publicity factor}. The hotness of a topic (or keyword) depends on how popular it is. People always find the hot topic for reading. 
\item[iii-]~~\textit{Impact factor}. In the world of scientific publications the citation is a very important factor. People often follow the citations of a scientific paper for finding the necessary information; 
\item[iv-]~~\textit{Randomness factor}. Randomness is very important factor in many real complex systems. Readers sometimes find something for reading quite randomly.
\end{itemize}
SciRecSys recommendation system is designed to be a search engine for scientific publications. We want to apply the rank, inference and similarity of keywords to solve three following problems: (i) Ranking papers matched to the given keyword; (ii) Recommending additional keywords related to a given keyword to help the reader navigating the corpus effectively; (iii) Suggesting papers for \textit{``Reading more''} function in the context the reader is reading a given topic.  
 
\section{Related works}\label{Sect:Related}
The similarity of keywords is often used as a crucial feature to reveal the relation between objects and many methods to measure it have been proposed. 
One baseline for similarity measures is using distance functions such as squared Euclidean distance, cosine similarity, Jaccard coefficient, Pearson’s correlation coefficient, and relative entropy. Anna Huang et al. \cite{huang2008similarity} has compared and analyzed the effectiveness of these measures in text clustering problem. She represent a document as an m-dimensional vector of the frequency of terms and uses a weighting scheme to reflect their importance through frequencies tf/idf. Singthongchai et al. \cite{singthongchai2013method} make keywords search more practical by calculating keyword similarity by combining Jaccard's, N-Gram and Vector Space. Probabilistic models for similarity measure has been studied in language speech \cite{dagan1999similarity,cha2007comprehensive}. Ido Dagan et al.  have proposed a bigram  similarity  model and used the  relative  entropy to compute the similarity of keywords \cite{dagan1999similarity}. Sung-Hyuk Cha has conducted a comprehensive survey on probability density functions for similarity measures \cite{cha2007comprehensive}. 
Ontology-based methods are also exploited to measure the similarity of keywords \cite{bollegala2007measuring,schickel2007oss,sanchez2012ontology}. Bollegala et al. \cite{bollegala2007measuring} use the Wordnet database - ontology of words to measure keywords' relatedness by extracting lexico-syntactic patterns that indicate various aspects of semantic similarity and modifying four popular co-occurrence measures, including Jaccard, Overlap (Simpson), Dice, and Pointwise mutual information (PMI). Vincent Schickel-Zuber and Boi Falting \cite{schickel2007oss} present a novel similarity measure for hierarchical ontologies called Ontology Structure based Similarity (OSS) that allows similarities to be asymmetric. Sánchez et al. \cite{sanchez2012ontology} presents an ontology-based method relying on the exploitation of taxonomic features available in an ontology. 

Markov Chain model which properties are studied in \cite{keener1993perron,mpagerank} is used for computing the similarity and ranking \cite{fouss2007random,le2014general}. Fouss et al.\cite{fouss2007random} use a stochastic random-walk model to compute similarities between nodes of a graph for recommendation. Vu et al. \cite{le2014general} introduce an N-star model, and demonstrate it in ranking conference and journal problems. Finally, Lops et al. \cite{lops2011content} do a thorough review on state-of-the-art and trends of content-based recommender systems.

\section{Backgrounds}\label{Sect:Backgrounds}
\subsection{Basic definitions}\label{Sect:Definitions}
Suppose $\mathcal{K}$, $\mathcal{P}$ are the sets of keywords and papers respectively. $p\in \mathcal{P}$ is a paper and $A\in \mathcal{K}$ is a keyword. $K(p) \subseteq \mathcal{K}$ is the set of keywords belonging to $p$. $P(A) = \{ q \in \mathcal{P}|A \in K(q)\}$ is the set of papers containing $A$. We assume that $K(p)\neq \emptyset \wedge P(A)\neq \emptyset$.  Finally, $C(p)\subseteq \mathcal{P} $ is the set of papers cited by $p$. In the case $p$ has no citing, we assume that $C(p)=\mathcal{P}$. It guarantees that $C(p)\neq \emptyset$.

A couple $(\mathcal{A},R)$ is called a \textit{ranking system} if: (i) $\mathcal{A} = \{a_1,\ldots,a_n\}$ is a finite set, and (ii) $R$ is a non-negative function on $\mathcal{A}$ (i.e., $R:\mathcal{A}\rightarrow[0,+\infty)$).
A \textit{ranking score} on $\mathcal{A}$ can be represented as $R = (R(a_1),\ldots,R(a_n))^T$. 
Furthermore, $(\mathcal{A},R)$ is \textit{normalized} if  $R^T$ is normalized ($\parallel R^T\parallel=1$).

Suppose $A$ and $B$ are two events. The \textit{inference} and \textit{similarity} of two events $A$ and $B$ are denoted by $I(A,B)$ and $S(A,B)$, respectively. They can be computed as follows:
\vspace*{-0.2cm}
\begin{equation}\label{eq1}
I(A,B)= \frac{Pr(A~and~B)}{Pr(A)} ~~~~~~~~~~~~~~ S(A,B)= \frac{Pr(A~and~B)}{Pr(A~or~B)}
\end{equation}
We have $S(A,B)\leq 1$, $I(A,B)\leq 1$. $S(A,B)$ is symmetric. $I(A,B)=Pr(B|A)$.     
\subsection{Relationships of keywords based on the occurrences}\label{Sect:Relationships}
Let us introduce two normalized ranking scores $R^{c}$, $R^{p}$  on set of keywords, $\mathcal{K}$, based on the occurrences. $R^{c}$ ($c$ stands for \textit{counting}) is based on counting the documents containing the given keyword. 
\vspace{-0.3cm}
\begin{equation}\label{eq3}
R^{c} (A) = \frac{|P(A)|}{\Sigma_{B\in \mathcal{K}} |P(B)|}~~~~~~~~~~( A\in \mathcal{K})
\vspace{-0.1cm}
\end{equation} 
$R^{p}$ ($p$ stands for \textit{probability}) is based on the probability of the occurrence of the given keyword in the papers. 
\vspace{-0.4cm}
\begin{equation}\label{eq4}
R^{p} (A) = \frac{1}{|\mathcal{P}|}\Sigma_{p\in P(A)}\frac{1}{|K(p)|} ~~~~~~~( A\in \mathcal{K})
\vspace{-0.2cm}
\end{equation}
$\frac{1}{|\mathcal{P}|}$ is the probability of choosing a paper from the corpus. $\frac{1}{|K(p)|}$ is the  probability of choosing keyword $A$ from the paper $p$ containing $|K(p)|$ keywords.

We propose following formulas for measuring the inference and similarity of two keywords $A$, $B$ based on the occurrences:
\vspace{-0.1cm}
\begin{equation}\label{eq5}
I^c(A,B)= \frac{|P(A)\cap P(B)|}{|P(A)|}~~~~~S^c(A,B)= \frac{|P(A)\cap P(B)|}{|P(A)\cup P(B)|} ~~~~( A,B\in \mathcal{K})
\vspace{-0.2cm}
\end{equation}
$P(A)\cap P(B)$ is the set of papers containing both keywords $A$ and $B$. $P(A)\cup P(B)$ is the set of papers containing  keywords $A$ or $B$.
\section{Relationships of keywords based on graph of keywords}\label{Sect:Our Approach}
\subsection{Markov Chain model of the reading process}\label{Sect:Markov}
We propose a Markov Chain model to simulate the reading process which can combine four factors: content, publicity, impact and randomness. We assume that the reader reads a topic (keyword) $A$ in some paper $p$ at any time ($A\in K(p)$). $\mathcal{S} = \{(A,p)\in \mathcal{K}\times \mathcal{P}| A \in K(p)\}$ is the set of states. From current state $\theta=(A,p)$, the reader will move to new state $\xi=(B,q)\in \mathcal{S}$ with  the conditional probability $Pr(\theta\rightarrow \xi)$ by applying 4 following actions:

\begin{itemize}
\item $A_1$. \textit{Same paper - some topic}. The reader is interested in current paper, and choose randomly a topic in the same paper with the probability equal to $\alpha_1$. Thus, $p=q$. 
\vspace{-0.1cm}
\begin{equation}\label{eq7}
Pr(\theta\rightarrow_{A_1}\xi) = if(p=q,\frac{\alpha_1}{|K(p)|},0)
\vspace{-0.1cm}
\end{equation}
\item $A_2$. \textit{Same topic - some paper}. The reader is interested in current topic, and choose randomly another paper which have the same topic with the probability equal to $\alpha_2$. Thus, $A=B$. 
\vspace{-0.2cm}
\begin{equation}\label{eq8}
Pr(\theta\rightarrow_{A_2}\xi) = if(A=B,\frac{\alpha_2}{|P(A)|},0)
\vspace{-0.1cm}
\end{equation}
\item $A_3$. \textit{Some topic - cited paper}. The reader is interested in some cited paper  with the probability equal to $\alpha_3$. First, he choose randomly a cited paper and then choose randomly a new topic belonging to the chosen paper. Thus, $q\in C(p)$. 
\vspace{-0.2cm}
\begin{equation}\label{eq9}
Pr(\theta\rightarrow_{A_3}\xi) = if(q\in C(p),\frac{\alpha_3}{|C(p)||K(q)|},0)
\vspace{-0.1cm}
\end{equation}
\item $A_4$. \textit{Some paper - some topic}. The reader stop reading the current paper and choose randomly a new paper with the probability equal to $\alpha_4$. 
\vspace{-0.2cm}
\begin{equation}\label{eq10}
Pr(\theta\rightarrow_{A_4}\xi) = \frac{\alpha_4}{|\mathcal{P}||K(q)|}
\vspace{-0.1cm}
\end{equation}
\end{itemize}
$\alpha_i>0$ are constants and $\Sigma^4_{i=1}{\alpha_i} =1 $.  From the assumptions, we have:
\vspace{-0.1cm}
$$Pr(\theta\rightarrow\xi) = \Sigma^4_{i=1} Pr(\theta\rightarrow_{A_i}\xi)$$ 
\vspace{-0.1cm}
The necessary condition of the Markov Chain model is that
total of the output conditional probability from any state is equal to 1. Here is the formula of the conditions:
$$O(\theta) = \Sigma_{\xi}  Pr(\theta\rightarrow\xi) = 1 $$
Our readers can check it by apply the formula \ref{eq7}, \ref{eq8}, \ref{eq9} and \ref{eq10}.
The probability of a reader in state $\theta=(A,p)$ is denoted by $Pr(\theta)$. We have:
\vspace{-0.2cm}
\begin{equation}\label{equaProb}
Pr(\theta)= \Sigma_{\xi\in \mathcal{S}} Pr(\xi)\times Pr(\xi\rightarrow\theta)   
\end{equation}
\vspace{-0.75cm}
\begin{proposition}
There exists a unique stationary score $\{Pr(\theta)\}_{\theta\in\mathcal{S}}$ satisfying (\ref{equaProb}).
\end{proposition}
\vspace{-0.25cm}
\begin{proof}
Since a state $\theta$ can jump to any state (and itself too) by apply action $A_4$, the Markov Chain is irreducible and aperiodic (see more \cite{keener1993perron,mpagerank}). The Perron-Frobenius theorem \cite{keener1993perron} states that there exists a unique stationary score $\{Pr(\theta)\}_{\theta\in\mathcal{S}}$.
\end{proof}
\vspace{-0.2cm}
$\{Pr(\theta)\}_{\theta\in\mathcal{S}}$ is determined by following algorithm:\\
{\small
\line(1,0){300}
~\\
{\bf Algorithm :} Computing  Stationary probability $\{Pr(\theta)\}_{\theta\in\mathcal{S}}$
~\\[-2mm]
\line(1,0){300}\\
\textbf{{\scriptsize 1.}~~begin}\\
\textbf{{\scriptsize 2.}~~~~~~~}$k=0$\\
\textbf{{\scriptsize 3.}~~~~~~~Foreach}  $~~~~\theta\in \mathcal{S}~~~~~~$ {\bf do} $~~~Pr(\theta)^{\texttt{\tiny(0)}}=\frac{1}{|\mathcal{S}|}$ \\
\textbf{{\scriptsize 4.}~}~~~~~~{\bf repeat}\\
\textbf{{\scriptsize 5.}~}~~~~~~~~~~~~$k = k + 1$  $~~~~stop = true$\\
\textbf{{\scriptsize 6.}~}~~~~~~~~~~~~{\bf Foreach}  $~~~~\theta\in \mathcal{S}~~~~~~$ {\bf do}\\
\textbf{{\scriptsize 7.}~}~~~~~~~~~~~~~~~~~$~Pr(\theta)^{\texttt{\tiny(k)}}=0$\\
\textbf{{\scriptsize 8.}~}~~~~~~~~~~~~~~~~~~~{\bf Foreach}  $~~~~\xi\in \mathcal{S}~~~~~~$ {\bf do}
 $~~~Pr(\theta)^{\texttt{\tiny(k)}} += Pr(\xi)^{\texttt{\tiny(k-1)}} Pr(\xi\rightarrow\theta)  $\\
\textbf{{\scriptsize 9.}~}~~~~~~~~~~~~~~~~~~~{\bf if}  $~~~~|Pr(\theta)^{\texttt{\tiny(k)}} - Pr(\theta)^{\texttt{\tiny(k)}}| > \epsilon~~~~$ {\bf then} $~~~~~~stop=false$\\ 
\textbf{{\scriptsize 10.}~}~~~~~~{\bf until} $stop$\\
\textbf{{\scriptsize 11.}~~~~~~~Foreach}  $~~~~\theta\in \mathcal{S}~~~~~~$ {\bf do} $Pr(\theta) = Pr^{\texttt{\tiny(k)}}(\theta)$ \\
\textbf{{\scriptsize 12.}~end}
~\\[-2mm]
\line(1,0){300}
}
\subsection{Ranking, Inference and Similarity of keywords}\label{Sect:NewRel}
The rank score of keyword $A$ based on the transition graph, $R^g(A)$ ($g$ stands for \textit{graph}), is equal to the probability of user reads keyword $A$. We have:
\begin{equation}\label{eq11}
R^g(A) = \Sigma_{p\in K(A)} Pr(\theta)~~~~~~~(\theta=(A,p)\in S)
\end{equation} 
Let $n_0$ be a positive integer. For each sequence $s=\theta_1\theta_2\ldots\theta_{n_0} \in \mathcal{S}^{n_0}$, let $Pr(s)$ is the probability of $s$ occurs in $\mathcal{S}^{n_0}$  generated by the Markov Chain model. $F\subseteq \mathcal{S}^{n_0}$, we denote $Pr(F)=\Sigma_{s\in F} Pr(s)$. For each keyword $A$, let 
$$P^g(A)= \{s=\theta_1\theta_2\ldots\theta_{n_0} \in \mathcal{S}^{n_0}| \exists i\in\{1,2,\ldots,n_0\},p\in \mathcal{P}: \theta_i=(A,p) \}$$  
The formulas for measuring the inference and similarity of two keywords $A$, $B$ based on the Markov Chain model:
\begin{equation}\label{eq12}
I^g(A,B)= \frac{Pr(P^g(A)\cap P^g(B))}{Pr(P^g(A))}~~~~S^g(A,B)= \frac{Pr(P^g(A)\cap P^g(B))}{Pr(P^g(A)\cup P^g(B))}
\end{equation}
We will apply Monte Carlo method for Markov Chain to compute the formulas (\ref{eq12}). First, we generate a large enough number of $n_0$-length sequences of states by our Markov Chain model. Then  we apply counting techniques for approximating the probabilities and computing the formulas.   
\section{SciRecSys - Recommendation System}\label{Sect:SciRecSys}

In this paper, we want to present three scenarios by using SciRecSys. 

\begin{description}
\item[Universal ranking vs. keyword based ranking]
The reader chooses a keyword $A$ to find some related papers belonging to $P(A)$. Question is ``What is the order for sorting $P(A)$?''. There are two ways for ranking: (i) All results are sorted by  only one \textit{universal ranking}, $\{R_u^g (p)\}_{p\in\mathcal{P}}$ or (ii) The order of the results depend on $A$ with the \textit{keyword based ranking}, $\{R_A^g (p)\}_{p\in P(A)}$. We propose $R_u^g (p)$ is equal to the probability of user reads $p$. Hence,
\vspace{-0.2cm}
\begin{equation}\label{eqRC1}
R_u^g (p)=\Sigma_{A\in K(p),\theta=(A,p)}Pr(\theta)
\vspace{-0.2cm}
\end{equation}
We propose $R_A^g (p)$ is equal to the probability of state $(A,p)$. Hence,
\vspace{-0.2cm}
\begin{equation}\label{eqRC2}
R_A^g (p)=Pr(\theta)~~~(\theta=(A,p),p\in P(A))
\vspace{-0.2cm}
\end{equation}
\item[Recommending keywords]
The user is in keyword (topic) $A$. What are the next recommended topics for following situations: (i) similar topics, $Sibling(A)$? (ii) more detail topics, $Child(A)$? (iii) more general topics, $Parent(A)$?  Let $m_{c}, ~ m_{p},~m_{s}$ be the parameters of the system. We denote:
$Child(A) = \{B\in \mathcal{K}| I^g(B,A)>m_{c}\}$;
$Parent(A) = \{B\in \mathcal{K}| I^g(A,B)>m_{f}\}$;
$Sibling(A) = \{B\in \mathcal{K}| S^g(A,B)>m_{b}\}$.
We remind our readers that keyword $A$ may have many parents or his parent can be his sibling or his child. 

\item[Recommending papers]
The user is in paper $p$. The user can choose the most interesting keyword $A \in K(p)$ to read more. What are the next recommended papers for him?  The system will request the user to refine recommended papers by choosing a keyword $A$ and then give the recommendation based on the conditional probability change from state $\theta=(A,p)$ to the papers by applying actions $A_2$ and $A_3$.  Suppose:
\vspace{-0.2cm}
\begin{equation}
R^g_{\theta}(q)=  \Sigma_{B\in K(q),\xi=(B,q)}(Pr(\theta\rightarrow_{A_2}\xi)+Pr(\theta\rightarrow_{A_3}\xi))
\vspace{-0.2cm}
\end{equation}
Finally, the recommended papers are chosen based on $R^g_{\theta}$ ranking function.  
\end{description}

\section{Experiments}\label{Sect:Experiments}
We collect data from DBLP\footnote{\url{http://dblp.uni-trier.de} accessed on December 2013} and Microsoft Academic Search\footnote{\url{http://academic.research.microsoft.com/} accessed on December 2013} (MAS) to conduct experiments. We choose three datasets for experiments from three different domains: (i) $D_1$ is the publications of ICRA - International Conference on Robotics and Automation; (ii) $D_2$  is the publications of ICDE - International Conference on Data Engineering (iii) $D_3$  is the publications of GI-Jahrestagung - Germany Conference in Computer Science. ICRA and ICDE conferences are one of the most famous and biggest ones in their area. They are chosen for rich publications and citations. GI-Jahrestagung is a smaller conference but its topic is quite various.

\begin{table}
\begin{center}
\caption{Experiment datasets}
\label{table:1}
    \begin{tabular}{c c c c c}
    \hline
    \textbf{Datasets} & \textbf{Paper No. } & \textbf{Keyword No. } & \textbf{Citation No.} & \textbf{State No.}\\ 
    \hline
    \hline
   ICRA & 9291 & 4676 & 125610 & 28736 \\
   ICDE & 3254 & 2755 & 99492 & 12365\\
   GI-Jahrestagung & 1335 & 1640 & 5 & 2932 \\
    \hline
    \end{tabular}
\end{center}
\vspace{-0.8cm}
\end{table}

\subsection{Experiments for rank scores} \label{Sect:ERank}
We do the experiments for three different ranking scores $R^c$, $R^p$ and $R^g$. The values on $ \{\alpha_i\}_{i=1}^4$ are chosen for testing different contexts. The rank scores are scaled with the same rate for the convenience. For each two ranking scores $i$ and $j$, we do examine: (i) the Spearman's rank correlation coefficient (denote $\rho_{ij}$) for monotone checking; (ii) the differences of values: $\Delta_{ij}(A)= R^{i}(A) - R^{j}(A)$, $\%\Delta_{ij}(A)= \frac{\Delta_{ij}(A)}{R^{j}(A)}$. Here are some interesting observations:
\begin{itemize}
\item \textit{The rank scores are monotone but quite different to each others}. It comes from that $\rho_{ij}$ are close to 1 and the average values of $\%\Delta_{ij}$ are very high. For instance with dataset $D1$, we have: $\rho_{cg}=0.97$, $\rho_{pg}=0.99$, $\rho_{cp}=0.97$. The average of $|\%\Delta_{cg}|$, $|\%\Delta_{pg}|$, $|\%\Delta_{cp}|$ are $47.13\%, 30.06\%$ and $51.76\%$ respectively. 
\begin{table}
\begin{center}
\vspace{-0.8cm}
\caption{Top 5 keywords most different using $R^{g}$ vs. $R^c$ on $D_1$.}
\label{table:d1}
\begin{minipage}{2in}
\begin{center}
{\scriptsize
\begin{tabular}{c c c c c c }
\hline
\textbf{Keyword}	& $R^{c}$ & $R^{p}$ & $R^{g}$ & $\Delta_{cg}$ & $\Delta_{cg}$\% \\
\hline
Mobile Robot & 924 & 1082 & 1221 & 297 & 32.16 \\
Robot Hand & 140 & 230 & 276 & 136	& 97.40 \\
Path Planning & 313	& 389 & 442	& 129 & 41.19 \\
Motion Planning	& 393 & 483	& 511 & 118	& 29.96 \\
Visual servoing	& 248 & 316	& 350 & 102	& 41.02 \\
\hline
\end{tabular}
\\
(a) Top 5 increasing values
}
\end{center}
\end{minipage}
  \qquad
\begin{minipage}{2in}
\begin{center}
{\scriptsize
\begin{tabular}{c c c c c c}
\hline
\textbf{Keyword}	& $R^{c}$ & $R^{p}$ & $R^{g}$ & $\Delta_{cg}$ & $\Delta_{cg}$\% \\
\hline
Indexing Terms & 419 & 228 & 251 & -168 & 40.19 \\
Satisfiability & 109 & 66 & 83 & -26 & 23.77 \\
Real Time & 422	& 366 & 397	& -25 & 6.02 \\
Extended Kalman Filter	& 70 & 37 & 45 & -25 & 35.04 \\
Computer Vision & 58 & 37 & 35 & -23 & 39.11 \\
\hline
\end{tabular}
\\
(b) Top 5 decreasing values
}
\end{center}
\end{minipage}
\end{center}
\vspace{-0.8cm}
\end{table}
\item \textit{$R^{g}$ reflects how hot a keyword is}. Let us see the result shown in  Tab.~\ref{table:d1}. All \textit{increasing} keywords are hot topics now and all \textit{decreasing} keywords seem not being currently hot topics for conferences on robotics and automation. Let us see the example of two keywords \textit{Robot Hand} and \textit{Satisfiability}. They have quite the same $R^c$ values but $R^g$ values are four times difference. The explanation is that the hot topics have rich citations which increase the $R^g$ values.
\item \textit{$R^{p}$ is more acceptable to $R^{g}$ than $R^{c}$}. Let us see Tab.~\ref{table:d1} again. For top 5 \textit{increasing} keywords, $R^{c}< R^{p}< R^{g}$.  For top 5 \textit{decrease} keywords, $R^{p}< R^{g}< R^{c}$. The explanation is that the probability occurrences reflect the world more exactly than the counting but they still neglect the latent information like the citation graphs in the data corpus. 
\end{itemize}

\subsection{Experiments for Inference and Similarity} \label{Sect:ESim}
We do experiments to compare ($I^{c}$, $S^c$) vs.  ($I^{g}$,$S^{g}$). For both inference and similarity, we find out some amazing results:
\vspace{-0.1cm}
\begin{itemize}
\vspace{-0.1cm}
\item \textit{$I^{g}$ \& $S^{g}$ exploits the citation network and the results are domain dependent}. Our approach seems to give ``hot" results. Let us see Tab. \ref{table:realtime}(a), which shows similar keywords to \textit{Real Time} over dataset $D_1$. $S^{g}$ generates ``hot" keyword \textit{Humanoid Robot} instead of \textit{Indexing Terms}. This phenomenon also occurs in dataset $D_2$. We assume the cause is $I^{g}$ \& $S^{g}$ exploit citation network and the paper--keyword relationships to generate more attractive keywords. Additionally, similar keywords to \textit{Real Time} are quite different between two datasets $D_1$ and $D_2$, so we can say that these methods are domain dependent.
\vspace{-0.01cm}
\item \textit{$I^{g}$ \& $S^{g}$ could help overcome the missing co-occurrence problem}. The traditional methods based on occurrences of keywords could not work when there is no co-occurrence in the dataset. For instance, Tab. \ref{table:sensornetwork}(b) shows that $S^{c}$ generates only one similar keyword to \textit{Sensor Network} in small dataset $D_3$. In contrast, $S^{g}$ could help us find more acceptable similar keywords.

\begin{table}
\vspace{-0.8cm}
\begin{center}
\caption{Top 5 similar keywords of \textit{``Real Time''} using $S^{c}$ vs. $S^g$.}
\label{table:realtime}
\begin{minipage}{2.2in}
\begin{center}
{\scriptsize
\begin{tabular}{c c c}
\hline
$D_1$ &  $S^{c}$ & $S^{g}$ \\
\hline
1 & Mobile Robot & Path Planning \\
2 & \textbf{Indexing Terms} & Mobile Robot \\
3 & Path Planning & Motion Planning \\
4 & Obstacle Avoidance & Obstacle Avoidance \\
5 & Motion Planning &  \textbf{Humanoid Robot} \\
\hline
\end{tabular}
\\
(a) $D_1$ in robotics domain, $|K|=4676$
}
\end{center}
\end{minipage}
 \qquad
\begin{minipage}{2.3in}
\begin{center}
{\scriptsize
\begin{tabular}{c c c}
\hline
$D_2$ & $S^{c}$ & $S^{g}$ \\
\hline
1 & Stream Processing & Data Stream\\
2 & Data Stream & Stream Processing\\
3 & Database System & Database System\\
4 & Sensor Network & Object Oriented \\
5 &  Query Processing & Data Model\\
\hline
\end{tabular}
\\
(b) $D_2$ in database domain, $|K|=2755$
}
\end{center}
\end{minipage}
\end{center}
\vspace{-1cm}
\end{table}

\begin{table}
\begin{center}
\caption{Top 5 similar keywords of \textit{``Sensor Network''} using $S^{c}$ vs. $S^g$.}
\label{table:sensornetwork}
\begin{minipage}{2.2in}
\begin{center}
{\scriptsize
\begin{tabular}{c c c}
\hline
$D_2$ & $S^{c}$ &  $S^{g}$ \\
\hline
1 & Query Processing & Query Processing\\
2 & Data Management & Database System\\
3 & Real Time  & Data Management\\
4 & Data Stream  & Data Stream\\
5 & Satisfiability & Query Evaluation\\
\hline
\end{tabular}
\\
(a) Top 5 on $D_2$, $|K|= 2755$.
}
\end{center}
\end{minipage}
 \qquad
\begin{minipage}{2.3in}
\begin{center}
{\scriptsize
\begin{tabular}{c c c}
\hline
$D_3$ & $S^{c}$ & $S^{g}$\\
\hline
1 & Distributed System & Mobile Device\\
2 & - & Web Service \\
3 &  - & Data Warehouse \\
4 & - & Open Source \\
5 & - & Middleware\\
\hline
\end{tabular}
\\
(b) Top 5 on $D_3$, $|K|= 1640$.
}
\end{center}
\end{minipage}
\end{center}
\vspace{-0.7cm}
\end{table}
\end{itemize}

We also find out some interesting observations about inference particularly:
\begin{itemize}
\item \textit{The inference between two keywords is asymmetric and $I^g$ exploits the citation network to generate a clear inference}. 

Unlike similarity between two keywords, inference is asymmetric, i.e., the inference between A and B differs from the one between B and A. Let us see Tab. \ref{table:assymetric}, the inference from \textit{Robot Hand} and \textit{Robot Arm} are almost different. More interestingly, \textit{Robot Hand} could infer \textit{Robot Arm} but \textit{Robot Arm} does not infer \textit{Robot Hand}, so there is a flow of inference here. Further examination shows that, \textit{High Speed} is repeated at top 1 on two inference lists generated by $I^c$. Whereas, \textit{High Speed} only appears on $I^g$'s \textit{Robot Arm} inference list, not on \textit{Robot Hand} inference list. So, we have a clear inference flow from \textit{Robot Hand} to \textit{Robot Arm} then to \textit{High Speed} with $I^g$, not with $I^c$. To explain this difference, we notice that when we infer from \textit{Robot Hand}, \textit{High Speed} shares more common publications than \textit{Robot Arm} , 4 vs. 2. However, \textit{Robot Hand} has more citations from \textit{Robot Arm} than \textit{High Speed}, 9 vs. 5. So we assume that our approach exploits the citation network, which is an asymmetric structure, to generate a clearer inference.
\end{itemize}
\begin{table}
\vspace{-0.8cm}
\begin{center}
\caption{Top 5 inference keywords over $D_1$ using $I^{c}$ vs. $I^g$.}
\label{table:assymetric}
\begin{minipage}{2in}
\begin{center}
{\scriptsize
\begin{tabular}{c c c}
\hline
$D_1$	& $B$ from $I^c(A,B)$ & $B$ from $I^g(A,B)$\\
\hline
1 & \textbf{High Speed} & \textbf{Robot Arm}  \\
2 & Control System & Force Control \\
3 & Robot Arm & Three Dimensional \\
4 & Force Control & Autonomous Robot \\
5 & Parallel Manipulator & Parallel Manipulator\\
\hline
\end{tabular}
\\
(a) $A=$ \textit{Robot Hand}.
}
\end{center}
\end{minipage}
 \qquad
\begin{minipage}{2in}
\begin{center}
{\scriptsize
\begin{tabular}{c c c}
\hline
$D_1$	& $B$ from $I^c(A,B)$ & $B$ from $I^g(A,B)$\\
\hline
1 & \textbf{High Speed} & \textbf{High Speed} \\
2 & Obstacle Avoidance & Parallel Manipulator \\
3 & Control System & Three Dimensional \\
4 & Configuration Space & Control System \\
5 &  Parallel Manipulator  & Autonomous Robot \\
\hline
\end{tabular}
\\
(b) $A=$ \textit{Robot Arm}.
}
\end{center}
\end{minipage}
\end{center}
\vspace{-0.8cm}
\end{table}

\subsection{SciRecSys Recommendation System} \label{Sect:ERS}

We compare universal ranking scores $R^{g}_u(p)$ and local ranking score $R^{g}_A(p)$ and conduct  experimental models for recommending keywords. We get some interesting results:
\begin{itemize}
\item \textit{Universal ranking prefers most popular papers, whereas local ranking prefers papers being not only popular but also focused on a small set of topics}. 

Let us see Tab.~\ref{table:top5recpaper} (a). Top recommended papers returned by \textit{Universal ranking} shows a large number of citing/cited papers. Whereas, top recommended papers returned by \textit{Local ranking} have less citing/cited papers but their keywords are more specific. The average number of keywords in each recommended papers of \textit{Local ranking} and \textit{Universal ranking} are $2.25$ and $6.75$, respectively, except the common top ranked paper with $ID=395393$. These numbers in $D_2$ are $3.00$ and $10.12$. We notice that the average number of keywords in one paper is around 2 and 3 for those two datasets.

\begin{table}
\vspace{-0.8cm}
\begin{center}
\caption{Top 5 recommended papers for the keyword \textit{"Real Time"} using $R^{g}_u$ vs. $R^{g}_A$ ($|K|$: keyword no., $|C_i|$: citing no., and  $|C_e|$: cited no.).}
\label{table:top5recpaper}
\begin{minipage}{2.2in}
\begin{center}
{\scriptsize
\begin{tabular}{| c | c c c c | c c c c |}
\hline
$D_1$ & \multicolumn{4}{c|}{\textbf{Local Ranking} $R^g_{A}(P)$} & \multicolumn{4}{c|}{\textbf{Universal Ranking} $R^g_u$} \\
\hline
No.	& PaperID & $|K|$ & $|C_i|$ &  $|C_e|$ & PaperID & $|K|$ & $|C_i|$ &  $|C_e|$ \\
\hline\hline
1 & 395393&10& 23&336&395393&10&23&336
\\
2 & 1749067 & 2 & 11 & 21 & 1653515	& 9 & 19 & 272 \\
3 & 1867302& 1 & 0 & 3 & 278613	 &6 	& 35	&141 
\\
4 & 1791141	& 2 & 4 & 14 & 1662450 & 4& 23& 171
\\
5 & 1662450 & 4 & 23 & 171 & 1451499 & 8 & 13 & 105
\\
\hline
\end{tabular}
}
(a) Top 5 on $D_1$, $|P($\textit{"Real Time"}$)| = 422$.
\end{center}
\end{minipage}
 \qquad
\begin{minipage}{2.3in}
\begin{center}
{\scriptsize
\begin{tabular}{ | c | c c c c | c c c c |}
\hline
$D_2$ & \multicolumn{4}{c|}{\textbf{Local Ranking} $R^g_{A}(P)$} & \multicolumn{4}{c|}{\textbf{Universal Ranking} $R^g_u$} \\
\hline
No.	& PaperID & $|K|$ & $|C_i|$ &  $|C_e|$ & PaperID & $|K|$ & $|C_i|$ &  $|C_e|$  \\
\hline\hline
1 & 2158744	& 1	& 12 & 26 & 2848 & 8 &	19 & 65
\\
2 & 1209177	& 2 & 34 & 1 & 3045031 & 12	& 22 & 12
\\
3 & 1142083 & 2& 10& 0 & 832819 & 9 & 32	& 69
\\
4 & 2848 & 8 & 19 & 65 & 3396132 & 17	& 6 & 0
\\
5 & 1340498 & 2 & 18 & 1 & 308381 & 10 & 20 & 22
\\
\hline
\end{tabular}
}
(b) Top 5 on $D_2$, $|P($\textit{"Real Time"}$)| = 49$.
\end{center}
\end{minipage}
\end{center}
\vspace{-0.9cm}
\end{table}
\item \textit{Keywords recommended with the inference relationship depend on both latent information and data domain}. 

Let us see Tab.~\ref{table:top5reckeyword}(a) for ICDE, conference specialized in database. Three inference methods \textit{Children, Sibling} and \textit{Parent} generate three lists of specialized keywords in database domain. On the other hand, for GI-Jahrestagung, small conference with quite various topics in Tab.~\ref{table:top5reckeyword}(b), the recommendation lists are quite diversity. We also notice that $D_2$ have much more citations information than $D_3$.

\item \textit{The missing co-occurrence problem can be overcome by using inference relationship}. 

As mention above, our approach using stationary probability graph to generate new states follow Markov Chain Monte Carlo method. This helps to create new states that have the co-occurrence of keywords. So, we agree that the recommended lists computed by inference relationship showed in Table \ref{table:top5reckeyword} are acceptable.   
\begin{table}
\vspace{-0.8cm}
\begin{center}
\caption{Top 5 recommended keywords for the keyword \textit{"Sensor Network"}.}
\label{table:top5reckeyword}
\begin{minipage}{2.3in}
\begin{center}
{\scriptsize
\begin{tabular}{c  c c c }
\hline
$D_2$ & Children & Sibling & Parent \\
\hline\hline
1 & Query Processing & Query Processing & Data Management\\
2 & Database System & Database System& Query Processing
\\
3 & Indexation & Data Management & Data Stream \\
4 & Data Stream & Data Stream & Query Evaluation\\
5 & Data Management & Query Evaluation & Multi Dimensional\\
\hline
\end{tabular}
}
(a) Top 5 on $D_2$, $|K| = 2755$.
\end{center}
\end{minipage}
\qquad
\begin{minipage}{2.2in}
\begin{center}
{\scriptsize
\begin{tabular}{c c c c }
\hline
$D_3$ & Children & Sibling & Parent \\
\hline\hline
1 & Open Source  & Mobile Device & Mobile Device\\
2 & Mobile Device & Web Service & Data Warehouse
\\
3 & Web Service & Data Warehouse  & Self Organization \\
4 & Middleware & Open Source & Distributed System\\
5 & Data Warehouse & Middleware & P2P\\
\hline
\end{tabular}
}
(b) Top 5 on $D_3$, $|K| = 1640$.
\end{center}
\end{minipage}
\end{center}
\vspace{-0.9cm}
\end{table}
\end{itemize}

\section{Conclusion and future works}\label{Sect:Conclusion}
We have introduced and studied a new approach on discovering various relationships among keywords from scientific publications. The proposed Markov Chain model combined four main factors of the problem: content, publicity, impact and randomness. The stationary probability of the states in the model helps us ranking and measuring the inference and similarity of keywords.

We have proposed SciRecSys recommendation system for navigating the scientific publications efficiently. By applying the relationships among keywords, we have suggested the solutions for the ranking results of a given keyword, related keywords and recommended papers. The experiments have shown that our methods can reflect how hot keyword is and overcome the missing co-occurrence problem. Moreover, our approach can exploit the latent information of citation network effectively.

As future work, we are planning to $i$) do experiment on big dataset to upgrade the quality of our ranking and measuring system, $ii$) study how to combine inference relationship by stationary probability graph with a given ranking systems, $iii$) investigate the time series in the keyword relationship and the trend prediction problem, and $iv$) apply model of recommendation systems in various problems, e.g., product recommendation, event recommendation, and so on.
\vspace{-0.2cm}

\end{document}